\documentclass{emulateapj}

\shorttitle{Mid-Infrared Fundamental Plane}
\shortauthors{Jun and Im}
\begin{document}
\title{The Mid-Infrared Fundamental Plane of Early-Type Galaxies}
\author{Hyunsung David Jun\altaffilmark{1} and Myungshin Im\altaffilmark{1,2}}
\altaffiltext{1}{Astronomy Program, Department of Physics and Astronomy, FPRD, Seoul National 
University, Seoul 151-742, Korea; hsjun@astro.snu.ac.kr, mim@astro.snu.ac.kr.}
\altaffiltext{2}{Infrared Processing and Analysis Center, California Institute of Technology, 
Pasadena, CA 91125}

\begin{abstract}
 Three observables of early-type galaxies - size ($r_{e}$), surface brightness ($I_{e}$), and 
velocity dispersion ($\sigma_{0}$) - form a tight planar correlation known as the fundamental plane 
(FP), which has provided great insights into the galaxy formation and the evolution processes.
However, the FP has been found to be tilted against the simple virial expectation, prompting debates 
on its origin. In order to investigate the contribution of systematic stellar population variation 
to the FP tilt, we study here the FP relations of early-type galaxies in mid-infrared (MIR) which 
may represent the stellar mass well. We examined the wavelength dependence of the FP coefficients, 
$a$ and $b$ in $\log\,r_{e}= a\,\log\,\sigma_{0} + b\,\log\,\langle I \rangle_{e} + c$, using a 
sample of 56 early-type galaxies for which visible ($V$-band), near-infrared ($K$-band), and MIR 
(Spitzer IRAC, 3.6--8.0$\,\mu$m) data are available. We find that the coefficient $a$ increases as a 
function of wavelength as $da/d\lambda=0.11\pm0.04\,\mu m^{-1}$, while the coefficient $b$ reaches 
the closest to -1 at 3.6--5.8$\,\mu$m. When applied to the visible FP coefficients derived from a 
larger sample of nearby early-type galaxies, we get the FP relation with $(a,b) \simeq $ 
(1.6--1.8,\,-0.9) at 3.6$\,\mu$m. Our result suggests that the stellar population effect can explain 
more than half of the FP tilt, closing the gap between the virial expectation and the optical FP\@. 
The reduction in the FP tilt is reflected in the dynamical mass-to-light ratio, $M_{dyn}/L$, 
dependence on $L$ which decreases toward 3.6--5.8$\,\mu$m, suggesting that the MIR light better 
represents mass than the shorter wavelengths.
\end{abstract}


\section{Introduction}
 In the search for correlations among physical parameters of early-type galaxies, it has been 
recognized that the effective radius ($r_e$), the effective mean surface brightness ($\langle I 
\rangle_{e}$), and the central velocity dispersion ($\sigma_{0}$) form a planar relation (in 
logarithmic space) known as the fundamental plane (hereafter FP; \citealt{dre87}; \citealt{djo87}), 
in the form of $r_{e} \propto \sigma_{0}^{a}\,\langle I \rangle_{e}^{b}$ where $a$ and $b$ are 
found to be $(a,b) \simeq $ (1.2--1.5,\,-0.8) at visible wavelengths (\citealt{jor96}; 
\citealt{ber03}). Under the assumption of structural homology and a constant mass-to-light ratio, 
the virial theorem implies that the FP coefficients should be $(a,b)$=(2,\,-1) -- the so-called 
`virial expectation'. The observed discrepancy, or tilt of the FP with respect to the virial 
expectation has prompted many discussions to explain its origin (see \citealt{don06} for a review of 
this field).

 One of the ideas is that the tilt is caused by the systematic variation in the stellar population 
as a function of physical parameters such as galaxy luminosity.
 \citet{pah98b} investigated this effect by constructing the FP in $K$-band,
which is supposedly a good tracer of the stellar mass less affected by age and dust.
 Meanwhile, \citet{sco98} examined the wavelength 
dependence on the FP coefficients, and concluded that some of the tilt is caused by the stellar 
population manifested by the color-magnitude relation. These studies found that the stellar 
population effect exists, but it can only partially explain the tilt of the FP\@.

 More recent investigations tackle the FP tilt problem using new methods such as gravitational 
lensing (\citealt{tre06}; \citealt{bol07}) or dynamical modeling (\citealt{pad04}; \citealt{cap06}). 
Such studies suggest that the FP tilt nearly disappears when the FP is constructed in the 
mass-domain. Their conclusion is that the tilt must arise from the variation in the central 
mass-to-light ratio (\citealt{rob06}), but it is not clear whether the variation is dominated by 
dark matter or by stars \citep{bol07}.

 In this Letter, we extend the FP analysis to wavelengths beyond $K$-band to further investigate 
the effect of stellar population on the tilt. By doing so we aim to provide the FP that possibly 
better represents stellar mass (see \S\,5), and to improve the constraints on different scenarios 
for the FP tilt. 

\section{The sample}
 Early-type galaxies were chosen from the sample of \citet{pah99}, which was used to study the FP of 
nearby early-type galaxies in visible and near-infrared (hereafter NIR) passbands. The sample has 
the velocity dispersion information necessary for constructing the FP\@. We then searched for 
mid-infrared (hereafter MIR)\footnote{We designate these wavelengths MIR to distinguish them from 
the $K$-band.} archival images for galaxies in the \citet{pah99} sample. For the MIR data, we used 
the Spitzer Space Telescope, Infrared Array Camera (hereafter IRAC; \citealt{faz04}) images, 
covering four wavelength channels at 3.6, 4.5, 5.8, and 8.0$\,\mu$m. The flux-calibrated, mosaiced 
IRAC images were retrieved from the Spitzer archive for these objects.

\begin{deluxetable}{ccccccc}
\tablecolumns{7}
\tabletypesize{\scriptsize}
\tablecaption{Photometric parameters of the sample}
\tablewidth{0.47\textwidth}
\tablehead{
\colhead{$\lambda$} & \colhead{$r_{e,min}$} & \colhead{$r_{e,avg}$} & \colhead{$r_{e,max}$} & 
\colhead{$M_{min}$} & \colhead{$M_{avg}$} & \colhead{$M_{max}$}
\\($\mu$m) & (\arcsec) & (\arcsec) & (\arcsec) & (mag) & (mag) & (mag)}
\startdata
0.55& 2.1 & 20.1 &  81.1 & -23.4 & -21.4 & -19.4\\
2.2 & 2.3 & 14.8 & 104.0 & -26.7 & -24.5 & -22.6\\ 
3.6 & 2.0 & 14.3 &  65.3 & -25.8 & -23.6 & -21.8\\
4.5 & 2.2 & 14.7 &  80.5 & -25.1 & -23.0 & -21.2\\
5.8 & 1.2 & 15.2 &  90.0 & -25.1 & -22.7 & -21.0\\
8.0 & 1.1 & 13.6 &  86.8 & -24.2 & -22.3 & -20.5
\enddata
\tablecomments{Effective radii and absolute magnitudes from \citealt{pah99} ($V$- and $K$-band) and 
our Spitzer IRAC analysis (3.6--8.0$\,\mu$m) are presented in minimum, average, and maximum values.}
\end{deluxetable}

The surface brightness fitting was performed for these matched galaxies, and the objects satisfying 
$r_{e} > 2\,\arcsec$ for three or more IRAC-bands were retained for the FP analysis. We imposed 
this size limit in order to work with a sample with reliable $r_{e}$ values (see \S\,3.1).  After 
removing a few galaxies (NGC1275, NGC4824, NGC4478, NGC6166) that show peculiar light profiles 
(multiple source, close to a bright galaxy or stars), we finally identified 56 galaxies with IRAC 
data in five clusters (A0426, A1656, A2199, A2634, and VIRGO) satisfying our selection criteria. We 
present a brief summary of the photometric information in Table 1. The exposure times for the IRAC 
data range from 72 to 1000 secs.

 The above selection of the sample may introduce a bias in the derived FP coefficients
(\citealt{sco98}). However, such a bias would not affect our derivation of the wavelength dependence 
of the FP coefficients, since the multi-wavelength FP coefficients will be derived from the same 
galaxies for which the same bias would apply.

\section{Analysis of the data}
\subsection{Surface Brightness Fitting}
 IRAF ELLIPSE was used to obtain surface brightness profiles of our IRAC sample galaxies. We 
restricted the fitting region to $a > 2\,$pixels (along the semi-major axis) and discarded regions 
with S/$N_{rms}<\,$1. During the fit, we held the center, and fixed the ellipticities and the 
position angles of isophotes to those at the effective radius in the 3.6$\,\mu$m band. In addition, 
3$\,\sigma$ clipping was applied to reject outliers such as foreground stars. To subtract the 
background, we used the values determined from the SExtractor \citep{ber96}. The adaptive background 
mesh sizes were varied between 16 to 96\,pixels, and the best mesh was chosen to be the one which 
flattened the growth curve at the largest isophote ($a \sim\,$3--6$\,a_{e}$).

 After the ELLIPSE photometry, we used the de Vaucouleurs r$^{1/4}$ law to fit the observed surface 
brightness profiles measured along the semi-major axis\footnote{We also tried the Sersic r$^{1/n}$
law but found no difference in the FP coefficients; we therefore kept the r$^{1/4}$ methodology.}. 
The fitting procedure yields the effective radius (in $\arcsec$) $r_{e}=\sqrt{(b/a)_{e}}\,a_{e}$ 
where $a_{e}$ is the effective semi major axis and $(b/a)_{e}$ is the axis ratio of the isophote at 
this position. We tested the reliability of our fitting procedure using the simulated, PSF-convolved 
galaxies, and found that the surface brightness fitting gives unbiased, reliable results when $r_{e} 
> 2\,\arcsec$. At the same time, we get the mean surface brightness within $r_{e}$ (in AB 
magnitudes) $\langle\mu\rangle_{e} = m_{1/2} + 2.5\,\log\,(\pi r_{e}^2) - 10\,\log\,(1+z) - 
A_{\lambda} - K(z)$ where $m_{1/2}$ is the magnitude of the total flux within the effective isophote 
defined by $a_{e}$ and $b_{e}$, while cosmological dimming, galactic extinction ($A_{\lambda}$, 
using the formula of \citealt{lau94}, and the extinction curve of \citealt{fit07}), and 
K-correction are taken into account. The K-correction is computed using the spectral energy 
distribution of a 13\,Gyr age, solar metallicity, and 0.1\,Gyr burst model from \citet{bru03}, 
assuming the Salpter initial mass function. The last observable, $\sigma_{0}$ is a kinematic 
parameter and is not expected to vary as a function of wavelength; we consequently use the same data 
used for the visible and NIR bands \citep{pah99}.

 In our analysis, angular sizes were converted into physical length units for the FP construction
by setting the distance to A1656 as 98.1\,Mpc and calibrating the distances to individual clusters, 
utilizing the NIR FP \citep{pah98a} as a distance ladder.

\subsection{Fitting of FP Coefficients} 
 We fitted the FP coefficients of the multi-waveband sample in the following manner using a variety 
of methods:
\begin{equation}
\log\,r_{e}= a\,\log\,\sigma_{0} + b\,\log\,\langle I \rangle_{e} + c,
\end{equation}
where $\langle \mu \rangle_{e}$ and $\langle I \rangle_{e}$ are related as $\langle \mu \rangle_{e} 
\propto -2.5\,\log\,\langle I\rangle_{e}$. For the input $r_{e}$ and $\langle I \rangle_{e}$, we use 
our SB-fit results for MIR (\S\,3.1), and those listed in \citet{pah99} for V- and K-bands. 
 We tried five different fitting methods: standard least-squares fit, the inverse least-squares fit, 
the bisector of the two, the least-squares fit to the orthogonal plane, and the least absolute 
deviation fit to the orthogonal plane. These methods are outlined below. 

 It is natural to think of doing the standard least-squares fit of $\log\,r_{e}$ (hereafter LSQ; 
\citealt{guz93}; \citealt{ber03}), but early FP work mainly took $\log\,\sigma_{0}$ at the ordinate 
(\citealt{dre87}; \citealt{djo87}; hereafter inverse LSQ) for their purposes. More recent work
prefers the least-squares fitting of $\log\,r_{e}$ by minimizing the variance orthogonal to the FP 
plane (hereafter orthogonal least-squares fit, or OLSQ; \citealt{ber03}) or the least absolute 
deviations orthogonal to the plane (hereafter orthogonal least absolute deviation fit, or OLAD; 
\citealt{jor96}; \citealt{pah98a}). The orthogonal fitting has an advantage over other methods, 
reducing the systematic error by treating the variables symmetrically \citep{iso90}. However, the 
orthogonal methods yield larger measurement errors than the LSQ methods, especially for small 
samples \citep{iso90}.

 Therefore, we also estimated the FP coefficients by taking the bisector, or the plane equidistant 
from the planes obtained through the standard LSQ and inverse LSQ (hereafter the LSQ bisector). 
1,000 Monte Carlo samplings of subsets of early-type galaxies in \citet{ber03} were performed to 
derive the FP coefficient errors on a sample of 50 early types to justify our approach. Through the 
sampling, we found the errors of the FP coefficients to be $(\sigma_{a},\sigma_{b})=(0.14,0.06)$, 
best reproduced with the LSQ bisector method, while the other orthogonal methods gave overestimated 
errors ($\ga$50\,\% for the coefficient $a$). Aside from the error estimates, all three symmetrized 
methods reproduce the FP coefficient $a$ of \citet{ber03} and the $K$-band early-type galaxy sample 
of \citet{pah98a} within 5\,\% agreement. On the other hand, the standard and inverse LSQ methods 
are found to have about minus and plus 20\,\% systematic biases in the coefficient $a$ estimates 
respectively in comparison to the symmetrized methods. Given these results, we adopted the FP 
coefficients with the LSQ bisector method as our base result.

\section{Results}

\begin{deluxetable}{ccccc}
\tablecolumns{5}
\tabletypesize{\scriptsize}
\tablecaption{Constructed Fundamental Planes at visible through MIR}
\tablewidth{0.47\textwidth}
\tablehead{
\colhead{$\lambda(\mu$m)} & \colhead{$a$} & \colhead{$b$} & \colhead{$c$} & \colhead{$r$} 
\\(1) & (2) & (3) & (4) & (5)}
\startdata
0.55& 1.23 $\pm$ 0.10 & -0.86 $\pm$ 0.04 &  -9.16 $\pm$ 0.40 &  0.96\\
2.2 & 1.42 $\pm$ 0.11 & -0.81 $\pm$ 0.05 &  -8.20 $\pm$ 0.41 &  0.95\\ 
3.6 & 1.55 $\pm$ 0.11 & -0.89 $\pm$ 0.04 &  -9.89 $\pm$ 0.39 &  0.96\\
4.5 & 1.47 $\pm$ 0.11 & -0.92 $\pm$ 0.04 & -10.16 $\pm$ 0.41 &  0.96\\
5.8 & 1.57 $\pm$ 0.13 & -0.92 $\pm$ 0.05 & -10.55 $\pm$ 0.50 &  0.95\\
8.0 & 1.55 $\pm$ 0.14 & -0.75 $\pm$ 0.05 &  -9.30 $\pm$ 0.60 &  0.93
\enddata
\tablecomments{Fundamental planes for the sample of 56 galaxies with the $r_{e} > 2\,\arcsec$ cut 
using the LSQ bisector method. Each column represents (1) wavelength in $\mu$m, (2)--(4) plane 
coefficients $a$, $b$, and $c$ with associated uncertainties, and (5) the linear correlation 
coefficient.}
\end{deluxetable}

 In Table 2, we list the FP coefficients with errors from 1,000 bootstrap resampling (unless 
obtained directly from known error estimates, e.g., LSQ methods) derived from the LSQ bisector 
method, for wavelengths of 0.55--8.0$\,\mu$m. We further plot the result of the FP fit in Figure 1, 
overlayed on the data points. We caution readers to focus less on the absolute values of $(a,b)$, 
but to focus more on the trend of the values with wavelengths or methods (see discussions at the end 
of this section and \S\,2). Gathering the outcomes, we are led to the wavelength-dependent nature of 
the FP coefficients, with $(a,b)$ values getting close to the virial expectation of (2,\,-1) as the 
wavelength increases. Such a tendency has been noted before (\citealt{pah98b}; \citealt{sco98}), but 
our result indicates that it extends to 3.6$\,\mu$m, and possibly beyond. When each cluster was 
analyzed separately, we also find the trend.

\begin{figure}
\centering
\plotone{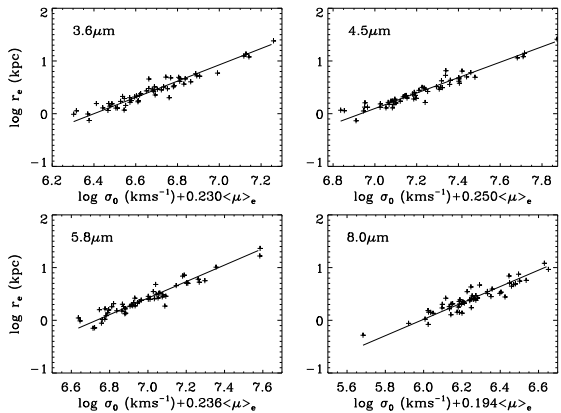}
\caption{Constructed fundamental planes projected in the direction of smallest scatter 
at 3.6, 4.5, 5.8, and 8.0$\,\mu$m, respectively. \label{fig1}}
\end{figure}

\begin{figure}
\center
\plotone{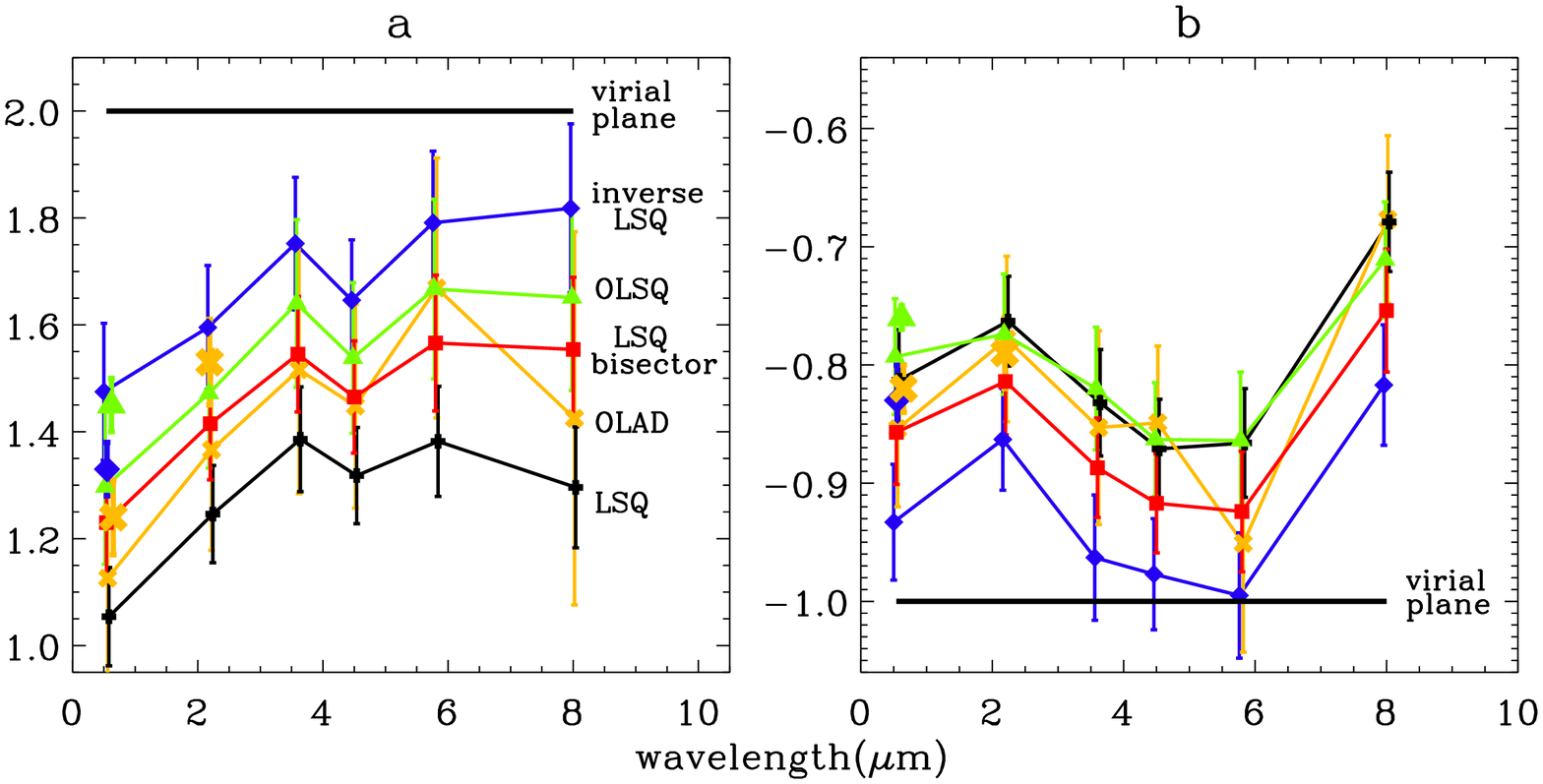}
\caption{Wavelength dependence of coefficients a and b; diamond, triangle, square, cross, and plus
sign symbols correspond to methodologies of inverse LSQ, OLSQ, LSQ bisector, OLAD, and ordinary LSQ 
respectively. Connected lines with spacing for identification are from our catalog, while solitary 
symbols are from other literature (\citealt{dre87}; \citealt{jor96}; \citealt{pah98a}; 
\citealt{ber03}) of nearby samples with N $\ga$ 100. Virial plane values assuming constant 
$M/L$ are $(a,b)$=(2,\,-1). 
\label{fig2}}
\end{figure}

 The wavelength dependence of the FP coefficients is further presented in Figure 2, where they are 
plotted using five different fitting methods (\S\,3.2). The tilt of the FP is maximally reduced 
toward the virial expectation at IRAC-bands, and the thickness of the FP is maintained thin for all 
but beyond 5.8$\,\mu$m. In terms of the methodology, we confirm the analysis of \S\,3.2 -- the three   
symmetrized fittings give coefficient $a$ values that are fairly consistent with each other 
(considering the sample size).

 To quantify the wavelength dependence, we model the change of coefficient $a$ as a linear function 
of wavelength by simultaneously fitting the OLSQ, OLAD, and the LSQ bisector results as follows:
\begin{equation}
da / d\lambda = 0.11\pm0.04\,\mu m^{-1}, 
\end{equation}
from the visible to 3.6$\,\mu$m (coefficient $a$ behaves flat afterward). This relation nicely 
explains the difference in coefficient $a$ of 0.05 in the SDSS g*- to z*-bands \citep{ber03}. 
Meanwhile, for $b$, the tendency is not as linear as that for $a$, but has a local maximum near the 
$K$-band, approaches closest to -1 at the IRAC 3.6--5.8$\,\mu$m bands, and increases again at 
8.0$\,\mu$m. We attribute this behavior at 8.0$\,\mu$m to the lower S/N, as well as the complexity 
in the 8.0$\,\mu$m emission which can be dominated by the dust emission in some cases 
(\citealt{bre06}; \citealt{ko07}). Indeed, the 8.0$\,\mu$m FP has the largest scatter among IRAC 
bands. The above result, together with the tendency of coefficient $b$ from Table 2, implies that 
the increase in coefficient $a$ and $b$ are ($\Delta a,\Delta b) \simeq $ (0.34,\,-0.03) from 
$V$-band to 3.6$\,\mu$m, and ($\Delta a,\Delta b) \simeq $ (0.15,\,-0.08) from $K$-band to 
3.6$\,\mu$m. If we use the FP coefficients from the references in Figure 2 as the base values on 
which to apply equation (2), we obtain $(a,b) \simeq $ (1.6--1.8,\,-0.9) at 3.6$\,\mu$m, which is 
quite close to the virial expectation. The implication of this result is discussed in the next 
section.
 
 Note that our coefficient $a$ in $K$-band, derived from a subsample of 56 early-types from 
\citet{pah98a} is smaller than the value derived from their full sample of 251 early-types by 
$\Delta a = -0.11$. The discrepancy should be mostly due to the limited sample size. More than half 
of our MIR galaxies belong to the Coma cluster (29 objects), and the Coma cluster galaxies in 
\citet{pah98a} show coefficient $a$ in the $K$-band ($a = 1.33$) smaller than the total sample 
result by $\Delta a = -0.20$, consistent with the results of \citet{mob99}. Apart from the 
wavelength dependence, our results seem to be skewed to the FP of the Coma cluster.
 
\section{Implications on the origin of the FP tilt}
 Recent studies suggest that the FP tilt originates mostly from a systematic variation in the  
mass-to-light ratio (\citealt{cap06}; \citealt{bol07}). However, the cause for the mass-to-light 
ratio variation is uncertain: it could be due to the stellar population, or the central dark matter 
fraction \citep{bol07}. Also, some studies suggest that the tilt is mostly explained by the 
non-homology related to the variation in the Sersic index n among early-type galaxies \citep{tru04}. 
Here, we discuss the implication of our result on these issues.

 First, we investigated which one of the parameters - size or luminosity - dominates the observed 
change in the FP coefficients with increasing wavelength. This was done by deriving the FP 
coefficients from the MIR sample by replacing (i) $r_{e}$'s or (ii) $\langle I \rangle_{e}$'s, with 
those from the shorter wavelength data (in our case the K-band). The result is presented in Figure 3 
(left), showing that the luminosity effect is the dominant factor up to 5.8$\,\mu$m. Interpretation 
at 8.0$\,\mu$m is difficult due to low S/N and dust emission. Our result suggests that the stellar 
population effect is significant going from K-band to IRAC-bands.

  Next, we examined to what extent the stellar population plays a role in the FP tilt through the 
dynamical mass-to-light ratio $M_{dyn}/L \propto r_{e}\sigma_{0}^{2}/L \propto 
\sigma_{0}^{2}/(r_{e}\langle I \rangle_{e})$ (e.g., \citealt{ber03}) variation calculated from the 
FP coefficients. If $M_{dyn}/L \propto L^{\beta}$, then $r_{e} \propto 
\sigma_{0}^{2/(1+2\beta)}\,\langle I \rangle_{e}^{-(1+\beta)/(1+2\beta)}$. The study of 
\citet{tru04} suggests $\beta \simeq 0.27$ based on the visible FP\@. Our result is that the FP 
coefficient reaches $a \simeq 1.6$--$1.8$ at 3.6$\,\mu$m (\S 4). In such a case, this relation gives 
$\beta \simeq $ 0.06--0.13, which enables us to explain more than half of the tilt in the visible 
FP\@. Moreover, the reduced tilt in the mass plane ($a_{MP} - a_{FP} = 0.27$: \citealt{bol07}) is 
consistent with our $\Delta a= 0.30 \pm 0.11$ from the I-band to the 3.6$\,\mu$m in equation (2), 
advocating that the $M_{dyn}/L$ variation is reduced by the regular light distributions in the 
MIR\@. As for the origin of the FP tilt, these results add another piece of evidence against the 
significance of non-homology (\citealt{pad04}; \citealt{cap06}; \citealt{bol07}), which predicts no 
change in the tilt with wavelength. 

\begin{figure}
\centering
\plotone{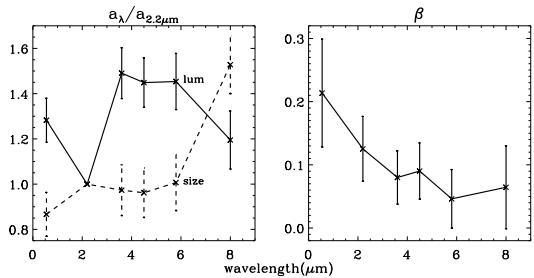}
\caption{\textit{Left}: The size vs luminosity test on the FP coefficient $a$. The solid line 
(labeled `lum') is for the set of coefficients computed by exchanging the $\langle I \rangle_{e}$ 
data of each wavelength with that from the K-band, while the dashed line (labeled `size') is the 
similar result by exchanging the $r_{e}$ data. \textit{Right}: The wavelength dependence of 
$M_{dyn}/L$ on $L$ as represented by the parameter $\beta$ of $M_{dyn}/L \propto L^{\beta}$ (see 
\S\,5).
\label{fig3}} 
\end{figure}

 We also derived the $\beta$ parameter by directly fitting the $M_{dyn}/L$. Figure 3 (right) 
demonstrates that the observed dependence of $M_{dyn}/L$ upon $L$ decreases and becomes flatter at 
IRAC-bands, just like the changes in $\beta$ derived from the FP coefficients. Combined with the 
fact that the change in the FP tilt with wavelength is dominated by the luminosity change, our 
$M_{dyn}/L$-fit result suggests that the rest-frame MIR luminosities at 3.6--5.8$\,\mu$m better 
represent the stellar mass than the shorter wavelengths, somewhat in agreement with \citet{tem08}, 
but not so with \citet{lac07}. Among many possibilities, a proper combination of the metallicity and 
the age variation can possibly reproduce the observed trend, and we plan to investigate as future 
work, the physical origin of the $M_{dyn}/L$ - $L$ relation as a function of wavelength.

\section{Summary}
 We studied the MIR fundamental plane relation of 56 early-type galaxies and derived the wavelength 
dependence of the FP coefficients. When the FP is expressed as $r_{e} \propto \sigma_{0}^{a}\,
\langle I\rangle_{e}^{b}$, we found that the exponent on $\sigma_{0}$, $a$, increases as a function 
of wavelength as $da/d\lambda = 0.11\pm0.04\,\mu m^{-1}$, while $b$ reaches closest to -1 without 
systematic variation with wavelength. When the FP coefficients of previous studies are adopted as 
the starting point to calculate the MIR FP coefficients, we found that $(a,b) \simeq $ 
(1.6--1.8,\,-0.9) which is closer to the virial expectation, and that the change is dominated by the 
luminosity change. Together with the reduced dependence of the $M_{dyn}/L$ on $L$ at MIR 
wavelengths, our outcomes suggest that the variation in the stellar population is responsible for a 
significant portion of the FP tilt, and that the rest-frame MIR better probes the stellar mass of 
low redshift early-type galaxies than the shorter wavelengths.

\acknowledgments
 This study was supported by a grant (R01-2007-000-20336-0) from the Basic Research Program of the 
Korea Science and Engineering Foundation, and by the Seoul Science Fellowship (HJ). We thank the 
referee for useful comments, and Youngmin Seo and Soonyoung Min for algorithmic and technical advice 
in data analysis.

\end{document}